\documentclass{aa}
\usepackage{epsf}
\begin{document}

   \thesaurus{06     
              (08.16.6;  
               08.13.1;  
	       08.18.1;  
               02.18.8)} 
   \title{Moments of inertia of relativistic magnetized stars}


   \author{K. Konno
          }


   \institute{Department of Physics, Hiroshima University,
              Higashi-Hiroshima 739-8526, Japan\\
              e-mail: konno@theo.phys.sci.hiroshima-u.ac.jp
             }

   \date{Received; Accepted}

   \maketitle

   \begin{abstract}

    We consider principal moments of inertia of axisymmetric,
    magnetically deformed stars in the context of general relativity.
    The general expression for the moment of inertia with
    respect to the symmetric axis is obtained.
    The numerical estimates are derived for several 
    polytropic stellar models.
    We find that the values of the principal moments of
    inertia are modified by a factor of 2 at most
    from Newtonian estimates.

   \end{abstract}

%

\section{Introduction}

Various features of pulsars have been disclosed with
growing observational data.
Most pulsars have stable pulse shapes, and spin down
steadily with typical time-scales of order 
several Myr (see e.g. M\'esz\'aros \cite{meszaros}). 
However, deviations from the linear
spin-down trends have also been observed on shorter 
time-scales in some pulsars. These deviations convey information
about the surroundings, internal structure and 
dynamics of neutron stars.
For example, the presence of glitches 
(see e.g.~Shemar \& Lyne \cite{glitch1}; Wang et al.~\cite{glitch2})
indicates the internal structure that two or more poorly 
coupled components coexist in the stars, while 
timing noise (see e.g.~D'Alessandro \cite{dmhd}) 
is believed to be due to 
random movements of the fluid in pulsars. 
Besides these short time-scale instabilities,
the spin-down features indicating the precession due to 
stellar deformation have also been reported.
Two anomalous X-ray pulsars (AXPs), 1E1048.1--5937
and 1E2259+586, were observed to indicate
irregular spin-down (see e.g.~Mereghetti \cite{mereghetti};
Baykal \& Swank \cite{bs}; Baykal et al.~\cite{bsss}; 
Oosterbroek et al.~\cite{opmi}).
AXPs are a subclass of X-ray pulsars with period in a very
narrow range (6--12s) and period derivatives in a range 
($10^{-12}-10^{-11}$ss$^{-1}$)
(see e.g.~Mereghetti \& Stella \cite{ms}), 
and those are most likely
candidates of magnetars (Duncan \& Thompson \cite{dt};
Thompson \& Duncan \cite{td1}; \cite{td2}; \cite{td3}) 
along with soft gamma repeaters (SGRs) 
(see e.g.~Kouveliotou et al.~\cite{sgr1}; 
Kouveliotou et al.~\cite{sgr2}).
As discussed by Melatos (\cite{melatos1}), the wobbles in 
their spin-down rates may be interpreted by an effect
called radiative precession (Melatos \cite{melatos2}),
which is related with an oscillating component of 
electromagnetic torque. 
Another precessing pulsar PSR B1828--11 
(Stairs et al.~\cite{sls}) was reported very recently.
This gives the very clear observation of free precession
of an isolated neutron star.
These deformed, precessing objects are analyzed by solving 
the Euler equation of motion in the form 
$I_{ij} d\Omega^{j} / dt - \varepsilon_{ijk} I^{jl} 
 \Omega^{k} \Omega_{l} = N_{i}$, where $I_{ij}$ is the 
inertia tensor, $\Omega^{i}$ is the angular velocity and
$N_{i}$ is the torque acting on the object.
When we take into account the electromagnetic torque
by a rotating magnetic dipole
(Davis \& Goldstein \cite{dg}; Goldreich \cite{goldreich}), 
the above-mentioned radiative precession can be found 
for magnetically deformed stars 
(Melatos \cite{melatos1}; \cite{melatos2}).
Furthermore, since magnetically deformed, rotating stars emit
gravitational waves,
we can consider the gravitational radiation 
reaction torque (Bertotti \& Anile \cite{ba};
Cutler \& Jones \cite{cj}). The gravitational backreaction
damps the wobbles on a time-scale proportional to
$[ \left( I_{z} - I_{x} \right)^2 / I_{x} ]^{-1}$
for wobbling, axisymmetric rigid bodies,
where $I_{i}$ is the principal moments of inertia.
Thus, the moments of inertia play a significant role 
in the analyses of pulsar precession. 
The estimates of the moments of inertia have been done
in the context of Newtonian gravity so far.
However, neutron stars are fully general relativistic
objects, and it is important to take into account
general relativistic effects. 
Therefore, we now discuss the principal moments of inertia
of magnetically deformed stars 
in the context of general relativity.

In Newtonian gravity, once we have the the mass distribution 
of an object $\rho ( \vec{x} )$, the moment of inertia
with respect to any axis can be calculated from 
$I=\int \rho ( \vec{x} ) \chi^2 d^{3}x$, where $\chi$ denotes the
length from the axis. In general relativity,
only for axisymmetric objects, the moment of inertia
with respect to the symmetric axis can be well defined.
For this definition, we need slow rotation of the 
objects. The slow rotation ensures that the angular
momentum $J$ is linearly related to the angular velocity 
$\Omega$, i.e.~$J=I \Omega$.
Here, $I$ defines the general relativistic version
of the moment of inertia.
The moments of inertia of relativistic, spherically
symmetric stars were discussed by Hartle (\cite{hartle})
and Chandrasekhar \& Millar (\cite{cm}).
The approximate expression valid for various realistic 
equations of state was also derived by 
Ravenhall \& Pethick (\cite{rp}).

In this paper, we discuss the principal moments of inertia
of magnetically deformed stars which are axisymmetric
with respect to the magnetic axis.
Deformation of relativistic magnetized stars was 
studied both in the numerical approach 
(Bonazzola et al.~\cite{bgsm}; Bocquet et al.~\cite{bbgn}) 
and in the analytic approach (Konno et al.~\cite{kok1}).
Based on the latter, analytic approach developed by
Konno et al.~(\cite{kok1}), we proceed with the discussions.
In the previous work (Konno et al.~\cite{kok1}), 
stellar magnetic fields were regarded as corrections to
non-rotating, spherical stars, and the magnetic deformation
of stars was formulated using a perturbative approach.
As mentioned above, we now take into account 
the slow rotation of the 
deformed stars on the symmetric axis to define 
the moment of inertia.
The formulation of this configuration is given in \S \ref{formul}.
The general expression for the principal moment of inertia
with respect to the symmetric axis 
is also derived. The numerical estimates are obtained 
for several stellar models in \S \ref{num}.
In \S \ref{oth}, we discuss the other principal moments of inertia.
Finally, we give conclusion in \S \ref{con}.
Throughout this paper, we use units in which
$c=G=1$.


\section{Magnetically deformed rotating stars}
\label{formul}

We consider a slowly rotating star which is subject to 
quadrupole deformation due to a dipole magnetic field.
As described by Konno et al.~(\cite{kok1}), 
magnetic fields can be treated as corrections to a 
background, spherically symmetric star, 
i.e.~$B \sim O( \varepsilon_{B} )$.
We now assume that the star slowly rotates on the magnetic axis
with a uniform angular velocity 
$\Omega \sim O( \varepsilon_{\Omega} )$,
which is also regarded as a perturbation.
In this paper, we take into account the rotational
corrections up to first order in $\varepsilon_{\Omega}$.
The metric describing such a star can be given by
\begin{eqnarray}
\label{metric}
 ds^2 & = & 
  - e^{\nu (r)} \left[ 1 + 2 \left( h_{0}(r) 
  + h_{2}(r) P_{2}(\cos \theta ) \right) \right] dt^2 \nonumber \\
 & &  + e^{\lambda (r)} \left[
  1 + \frac{2e^{\lambda (r)}}{r} \left( m_{0}(r) + 
  m_{2}(r) P_{2}(\cos \theta ) \right)\right] dr^2 \nonumber \\
 & & + r^2 \left( 1 + 2 k_{2}(r) P_{2} (\cos \theta ) \right) d\theta^2
     \nonumber \\
 & & + r^2 \sin^2 \theta \left( 1 + 2 k_{2}(r) 
     P_{2}(\cos \theta ) \right) \nonumber \\
 & & \quad \times  \left[ d\phi 
  - \left( W_{1}(r) - \frac{W_{3}(r)}{\sin \theta}
  \frac{dP_{3}(\cos \theta )}{d\theta} \right) dt \right]^2 ,
\end{eqnarray}
where $P_{l}$ is the Legendre polynomial of order $l$, 
and $\nu$ and $\lambda$ describe a background star.
The corrections $h_{0}$, $h_{2}$, $m_{0}$, $m_{2}$
and $k_{2}$ of second order in $\varepsilon_{B}$ 
correspond to deviation from the spherical shape.
If we replace $W_{1}$ and $W_{3}$ with zero in Eq.~(\ref{metric}),
this expression reduces to the metric already
investigated by Konno et al.~(\cite{kok1}).
The newly appeared functions $W_{1}$ and $W_{3}$
lead to frame dragging due to the 
magnetically deformed, rotating star. 
In order to involve the effect of deformation,
let these functions include corrections up to order 
$\varepsilon_{\Omega} \varepsilon_{B}^2$.
Showing order explicitly, we can write down 
$W_{1}$ and $W_{3}$ in the forms
\begin{eqnarray}
 W_{1} & = & \omega + W_{1}^{(2)} , \\
 W_{3} & = & W_{3}^{(2)} ,
\end{eqnarray}
where $\omega \sim O(\varepsilon_{\Omega})$ and
$( W_{1}^{(2)}, W_{3}^{(2)}) \sim 
O( \varepsilon_{\Omega} \varepsilon_{B}^2)$.
The form of the terms including $W_{1}$ and $W_{3}$ in
Eq.~(\ref{metric}) corresponds to quadrupole
deformation of the star.
These functions are very analogous to the 
rotational corrections
up to third order in $\varepsilon_{\Omega}$
(Quintana \cite{quin}; Kojima \& Hosonuma \cite{kh}).

The stress-energy tensor of the star which is
composed of the perfect fluid endowed with a 
dipole magnetic field has the form
\begin{eqnarray}
\label{s-e-t}
 T^{\mu}_{\ \nu} 
 & = & \left( \rho + p \right) u^{\mu} u_{\nu}
  + p \delta^{\mu}_{\ \nu} \nonumber \\
 & & + \frac{1}{4\pi} \left( F^{\mu \lambda}
  F_{\nu \lambda} - \frac{1}{4} F^{\sigma \lambda} F_{\sigma \lambda} 
  \delta^{\mu}_{\ \nu}\right) ,
\end{eqnarray}
where the density $\rho$ and the pressure $p$ are expanded as
\begin{eqnarray}
 \rho & = & \rho_{0}(r) + \left( \rho_{20}(r) +
   \rho_{22}(r) P_{2} \right) , \\
 p & = & p_{0}(r) + \left( p_{20}(r) +
   p_{22}(r) P_{2} \right) .
\end{eqnarray} 
Here, $\rho_{20}$, $\rho_{22}$, $p_{20}$ and $p_{22}$
are second-order quantities in $\varepsilon_{B}$.
The non-vanishing components of the 4-velocity $u^{\mu}$
are given by 
\begin{equation}
  u^{t} = e^{-\frac{\nu}{2}} \left[
  1-\left( h_{0} +h_{2} P_{2} \right) \right] , \quad 
  u^{\phi} = \Omega u^{t} .
\end{equation}
Furthermore, the Faraday tensor $F_{\mu \nu}$ can be
given by the 4-potential $A_{\mu}$ written as
(Muslimov \& Tsygan \cite{mt}; Konno \& Kojima \cite{kk})
\begin{equation}
 A_{\mu} = \left( a_{t0}(r) + a_{t2}(r) P_{2} , 0, 0, 
  a_{\phi}(r) \sin \theta \frac{dP_{1}}{d\theta} \right) ,
\end{equation}
where  
the $\phi$-component corresponds to a dipole
magnetic field, 
and the $t$-component denotes
the electric field induced by stellar rotation. 
Therefore, we have $a_{\phi} \sim O(\varepsilon_{B})$
and $(a_{t0}, a_{t2}) \sim O( \varepsilon_{\Omega} \varepsilon_{B})$.

The above-mentioned functions except $W_{1}$ and $W_{3}$ were
investigated in detail by Konno et al.~(\cite{kok1}) and
Konno \& Kojima (\cite{kk}) 
(see Appendix \ref{app} for the brief summary).
The differential equations which $W_{1}$ and $W_{3}$ obey can be 
obtained from the $(t\phi)$-component of the Einstein equation,
\begin{equation}
\label{de1}
 \frac{1}{r^4} \frac{d}{dr} \left[
   r^4 j \frac{d \overline{W}_{1}}{dr} \right] + \frac{4}{r} 
   \frac{dj}{dr} \overline{W}_{1}
 = - j \left( S_{1} - \frac{S_{2}}{5} \right) , 
\end{equation}
\begin{equation}
\label{de2}
 \frac{1}{r^4} \frac{d}{dr} \left[ r^4 j 
   \frac{d W_{3}}{dr} \right] + \left( \frac{4}{r} 
   \frac{dj}{dr} - \frac{10}{r^2} e^{\frac{\lambda-\nu}{2}} 
   \right) W_{3}
 = j \frac{S_{2}}{5} ,
\end{equation}
where $j$ and $\overline{W}_{1}$ are defined, respectively, as
\begin{eqnarray}
 j & = & e^{-\frac{\nu + \lambda}{2}} ,
\end{eqnarray}
\begin{eqnarray}
 \overline{W}_{1} & = & \Omega - W_{1} = \varpi - W_{1}^{(2)} 
  \quad (\varpi = \Omega - \omega ).
\end{eqnarray}
These differential equations are solved order by order.
The differential equation for $\varpi$ of order $\varepsilon_{\Omega}$,
in which both $S_{1}$ and $S_{2}$ vanish, was solved
by Hartle (\cite{hartle}).
We are now interested in the differential equations of order 
$\varepsilon_{\Omega} \varepsilon_{B}^2$.
The functions $S_{1}$ and $S_{2}$ appearing in the 
source terms in Eqs.~(\ref{de1}) and (\ref{de2}),
in general, include $W_{1}$ and $W_{3}$.
However, when we consider $S_{1}$ and $S_{2}$ 
up to the order of our interest, these functions include
$\Omega$ and $\omega$ only as shown below.
The source terms of order $\varepsilon_{\Omega} \varepsilon_{B}^2$ 
are given by
\begin{eqnarray}
\label{s1}
 S_{1} & = & \varpi' \left( - h_{0} - \frac{e^{\lambda}}{r}
   m_{0} \right)' - \frac{4}{r^2} e^{\lambda} \left( 
   \nu' + \lambda' \right) \varpi m_{0}
   \nonumber \\ 
 & & - 16 \pi  e^{\lambda} \varpi 
   \left( \rho_{20} + p_{20} \right) 
   + \frac{16 e^{\lambda}}{3 r^4} a_{\phi}^2 \omega
   \nonumber  \\
 & & + \frac{8}{3r^2} \left( a_{\phi}'\right)^2 \omega
   + \frac{8 e^{\lambda}}{r^4} a_{\phi} a_{t2} 
   - \frac{4}{r^2} a_{\phi}' a_{t0}' , \\
\label{s2}
 S_{2} & = & \varpi' \left( 4k_{2} - h_{2} - \frac{e^{\lambda}}{r}
   m_{2} \right)' - \frac{4}{r^2} e^{\lambda} \left( 
   \nu' + \lambda' \right) \varpi m_{2}
   \nonumber \\ 
 & &  - 16 \pi  e^{\lambda} \varpi 
   \left( \rho_{22} + p_{22} \right) 
   + \frac{32 e^{\lambda}}{3 r^4} a_{\phi}^2 \omega
   \nonumber  \\
 & & - \frac{8}{3r^2} \left( a_{\phi}'\right)^2 \omega
   + \frac{16 e^{\lambda}}{r^4} a_{\phi} a_{t2} 
   - \frac{4}{r^2} a_{\phi}' a_{t2}' ,
\end{eqnarray}
where the prime here denotes differentiation with respect to $r$.
These functions can be calculated by using  
the results for the density and pressure 
$(\rho_{20},\rho_{22}, p_{20}, p_{22})$,
the metric functions $(h_{0}, h_{2},
m_{0}, m_{2}, k_{2})$ and the potential functions 
$(a_{\phi}, a_{t0}, a_{t2})$, 
which were already calculated by Konno et al.~(\cite{kok1})
and Konno \& Kojima (\cite{kk}).

The differential equations (\ref{de1}) and (\ref{de2})
can be solved numerically by imposing 
boundary and junction conditions. 
The boundary conditions are summarized as
\begin{equation}
 W_{1} \rightarrow \mbox{const}, \quad
 W_{3} \rightarrow 0  \quad 
 \mbox{as} \quad r \rightarrow 0 ,
\end{equation}
\begin{equation}
\label{eq.bc.infty}
 W_{1}, W_{3} \rightarrow \frac{1}{r^{\beta}} 
 \left( \beta \ge 3 \right) \quad  
 \mbox{as} \quad r \rightarrow \infty .
\end{equation}
Furthermore, we impose the junction condition given by
(O'Brien \& Synge \cite{os}; 
see also Eqs.~(\ref{junc-1}) and (\ref{junc-2}))
\begin{eqnarray}
 \left. W_{a} \right|_{+ \xi^{(0)}} 
 & = & \left. W_{a} \right|_{- \xi^{(0)}} ,\\
 \left. W'_{a} \right|_{+ \xi^{(0)}} 
 & = & \left. W'_{a} \right|_{- \xi^{(0)}} ,
\end{eqnarray}
where $a$ takes 1 or 3, and $\xi^{(0)}$ denotes
the surface of the background star.

Before deriving numerical solutions, 
it is worthwhile to investigate 
the behavior of $W_{1}$ and $W_{3}$ at large $r$ in detail 
using Eqs.~(\ref{de1}), (\ref{de2}), (\ref{s1}) and (\ref{s2}),
because this inspection gives the expressions for
the angular momentum and the moment of inertia
of the star.

At large $r$, we have
\begin{equation}
 \lambda , \nu \rightarrow 0 , \quad 
 \mbox{i.e.} \quad j \rightarrow 1 .
\end{equation}
From Eqs.~(\ref{s1}) and (\ref{s2}), we derive
the asymptotic behaviors
\begin{equation}
 S_{1} \propto \frac{1}{r^8}, \quad 
 S_{2} \propto \frac{1}{r^7} .
\end{equation}
Using these expressions, from Eqs.~(\ref{de1}), (\ref{de2})
and (\ref{eq.bc.infty}), we obtain
\begin{equation}
 W_{1}^{(2)} \propto \frac{1}{r^3}, \quad
 W_{3}^{(2)} \propto \frac{1}{r^5} .
\end{equation}
Hence, we can put
\begin{equation}
 \label{afw1}
 \overline{W}_{1} = \Omega - \frac{2J}{r^3} - W_{1p}^{(2)} ,
\end{equation}
where $J$ is the angular momentum of the deformed star, and
$W_{1p}^{(2)} \sim O(1/r^6)$ is a function of order 
$\varepsilon_{\Omega} \varepsilon_{B}^2$. 
We now integrate Eq.~(\ref{de1}) after
multiplying both sides by $r^4$.
Using the above expression
(\ref{afw1}), we can obtain the general expression
for the principal moment of inertia
\begin{eqnarray}
\label{m-inertia}
 I_{z} & = & \frac{J}{\Omega} \nonumber \\
  & = & - \frac{2}{3} \int_{0}^{R} r^3
   \frac{dj}{dr} \frac{\varpi}{\Omega} dr
   - \frac{R^4}{6\Omega} 
   \left( \frac{dW_{1p}^{(2)}}{dr} \right)_{R} \nonumber \\
 & & + \frac{2}{3} \int^{R}_{0} r^3 
   \frac{dj}{dr} \frac{W_{1}^{(2)}}{\Omega} dr \nonumber \\
 & &   - \frac{1}{6 \Omega} \int^{R}_{0} r^4 j \left( S_{1} - 
   \frac{S_{2}}{5} \right) dr ,
\end{eqnarray}
where $R$ is the radius of the background star.
The moment of inertia of the background, spherically
symmetric star $I^{(0)}$ is given by 
(Hartle \cite{hartle}; Ravenhall \& Pethick \cite{rp})
\begin{equation}
 I^{(0)} = - \frac{2}{3} \int_{0}^{R} r^3
   \frac{dj}{dr} \frac{\varpi}{\Omega} dr .
\end{equation}
Therefore, we obtain
\begin{eqnarray}
\label{mci}
 I_{z}^{(2)} 
  & = & - \frac{R^4}{6\Omega} 
   \left( \frac{dW_{1p}^{(2)}}{dr} \right)_{R}
   + \frac{2}{3} \int^{R}_{0} r^3 
   \frac{dj}{dr} \frac{W_{1}^{(2)}}{\Omega} dr \nonumber \\
 & & - \frac{1}{6 \Omega} \int^{R}_{0} r^4 j \left( S_{1} - 
   \frac{S_{2}}{5} \right) dr ,
\end{eqnarray}
where we have used the decomposition
$I_{z} = I^{(0)} + I_{z}^{(2)}$.
The magnetic modification of the principal moment 
of inertia $I_{z}^{(2)}$ is the second order quantity 
in $\varepsilon_{B}$. The numerical estimates of 
$I_{z}^{(2)}$ for several stellar models
are given in the next section.

%

\section{Numerical estimates of the moment of inertia}
\label{num}

\begin{figure}
  \vspace{0cm}
  \hspace{0cm}\epsfxsize=8.8cm \epsfbox{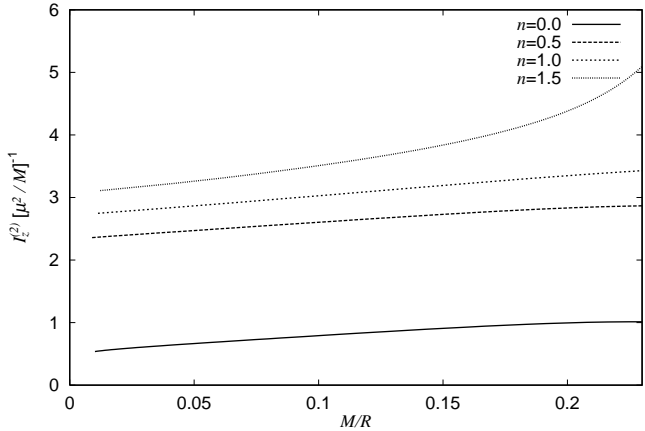}
  \vspace{0cm}
  \caption{The magnetic correction of the principal moment
      of inertia $I^{(2)}_{z}$ plotted as a function of $M/R$.
      The values are normalized by $\mu^2 / M$,
      and the polytropic index is denoted by $n$.}
  \label{figiz}
\end{figure}

Using the boundary and junction conditions mentioned 
in the last section, we can derive the numerical 
solutions for $W_{1}$ and $W_{3}$ for any stellar models.
The numerical results for the magnetic correction
of the principal moment of inertia $I^{(2)}_{z}$ can
also be obtained by using Eq.~(\ref{mci}).
We now show the results of $I^{(2)}_{z}$ obtained
for several polytropic stellar models.
In these calculations, we adopted a different condition
from that of the previous study (Konno et al.~\cite{kok1}),
in which sequences with constant central density were
investigated.
In the current case, in order to clarify the correspondence
between non-magnetized and magnetized stars having
same mass, we use the condition in which the total mass of the star
does not change through the perturbative approach.
This is also because the moment of inertia 
can be modified significantly by the mass shift
rather than the magnetic deformation.
In our formulation, this condition is accomplished
by imposing the boundary condition that $m_{0}$
vanishes at infinity.

Figure \ref{figiz} displays the magnetic correction
$I^{(2)}_{z}$ as a function of the general-relativistic factor $M/R$.
The values are normalized by the typical value $\mu^2 / M$.
As a simple example, we now discuss quadrupole deformation 
of a fluid body in the Newtonian limit.
If the deformed body has constant density, which
corresponds to $n=0$, then we can derive the result
$I^{(2)}_{z} = 10 \mu^2 / (3M)$
from the estimate of ellipticity $25 \mu^2 / (2M^2 R^2)$
(Ferraro \cite{ferraro}; Konno et al.~\cite{kok2}).
However, as seen in Fig.~\ref{figiz}, our numerical result 
for $n=0$ is different from this simple estimate. 
This is because the perturbed star does not have constant 
effective density by the added perturbation, 
even though we assume the background star with constant density.
This reason can be understood by seeing the differential 
equation for $m_{0}$, i.e.~Eq.~(\ref{de-m0}).
The derivative of $m_{0}$ is related to the effective density
including electromagnetic energy.
In the case of $n=0$, although the first term on the 
right-hand side in Eq.~(\ref{de-m0}) vanishes, the remaining
terms do not vanish and are non-trivial functions.
It follows that the effective density is not a constant.
Thus, our results include the inertia of electromagnetic fields
as well as mass.
Therefore, our result shown in Fig.~\ref{figiz}
cannot be compared with the above simple estimate.

Concentrating on the general relativistic effects on
the magnetic correction $I_{z}^{(2)}$, we can find 
from Fig.~\ref{figiz} that the values of $I_{z}^{(2)}$
for each stellar model become large with the
general relativistic factor $M/R$. The increments 
are 50\% at most. However, the rates of increase may be 
neglected except for the case of $n=1.5$.

\section{The other components of moments of inertia}
\label{oth}

\begin{figure}
  \vspace{0cm}
  \hspace{0cm}\epsfxsize=8.8cm \epsfbox{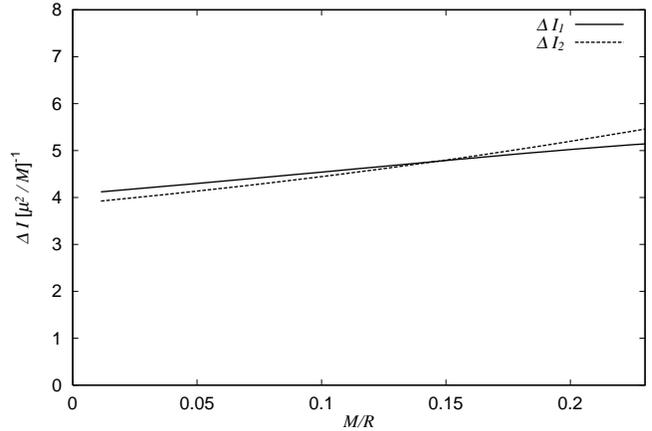}
  \vspace{0cm}
  \caption{Comparison between $\Delta I_{1}$ and $\Delta I_{2}$
      for $n=1$. $\Delta I_{1}$ and $\Delta I_{2}$ are
      normalized by $\mu^2 / M$ and plotted as a function of $M/R$.}
  \label{figid}
\end{figure}

Next, we discuss the other principal moments of inertia, 
i.e.~$I_{x}^{(2)}$ and $I_{y}^{(2)}$. 
In the context of general relativity, 
the definition of these quantities is associated with a 
difficulty in the concept.
As seen from Eqs.~(\ref{afw1}) and (\ref{m-inertia}), 
the notion of moments of inertia is related with
quantities at infinity (see also Geroch \cite{geroch}; 
Hansen \cite{hansen}).
If we consider the rotation of the star on the $x$-axis or
the $y$-axis, it produces the radiation of electromagnetic and 
gravitational fields.
Thus, the exterior space-time is not stationary, but radiative.
There is no rigorous way to define the moments of inertia
in such a radiative space-time. The concept
of $I_{x}^{(2)}$ and $I_{y}^{(2)}$ itself
may be meaningless to a considerable extent.

Therefore, only approximate expressions are available.
It would be useful to extend Newtonian expressions,
since it seems that most of Newtonian features 
still survive for neutron stars specified by 
the general relativistic factor $M/R \sim 0.2$.
We cannot prove the validity mathematically, but
will test some empirical relations.
For this purpose, let us recall relations using the principal
moments of inertia and the ellipticity which hold for an
incompressible fluid in the context of Newtonian gravity.
These are helpful in seeking the other principal 
moments of inertia.
First, if the fluid body is subject to
quadrupole deformation, then the relation between 
the principal moments of inertia exists
\begin{equation}
\label{q-rel}
 I_{x}^{(2)} = I_{y}^{(2)} = - \frac{1}{2} I_{z}^{(2)} .
\end{equation}
Using this relation, the difference between the 
principal moments of inertia $I_{z}^{(2)}$ and $I_{x}^{(2)}$
is given by
\begin{equation}
\label{i1}
 \Delta I_{1} = I_{z}^{(2)} - I_{x}^{(2)} = \frac{3}{2} I_{z}^{(2)} .
\end{equation}
In the different way, this difference can be expressed also by
using the ellipticity, which is defined by the difference
between the equatorial radius and the polar radius
(see e.g. Chandrasekhar \& Miller \cite{cm};
Konno et al.~\cite{kok1}; Konno et al.\cite{kok2}),
\begin{equation}
\label{i2}
 \Delta I_{2} = I^{(0)} \times (\mbox{ellipticity}).
\end{equation}
Thus, the difference is derived by the two methods.
There is no reason that the two expressions coincide,
but empirically the agreement is good.
Note that these expressions (\ref{i1}) and (\ref{i2})
coincide exactly in this case.

If we assume that the relation (\ref{q-rel}) is
applicable to general relativistic cases, it seems
that we should simply apply the relation (\ref{q-rel})
to the results shown in Fig.~\ref{figiz} and derive
$I_{x}^{(2)}$ and $I_{y}^{(2)}$.
However, in order to utilize this relation, the magnetic 
corrections must be purely quadrupole contributions.
As seen from the metric (\ref{metric}), the magnetic deformation
includes monopole parts as well as quadrupole parts.
Thus, our results in Fig.~\ref{figiz} also include monopole parts.
Hence, we have to subtract monopole contributions
from $I^{(2)}_{z}$ to derive $I_{x}^{(2)}$ and $I_{y}^{(2)}$. 
It is, nevertheless, not so easy to 
extract monopole contributions accurately from 
these magnetically deformed stars.

\begin{figure}
  \vspace{0cm}
  \hspace{0cm}\epsfxsize=8.8cm \epsfbox{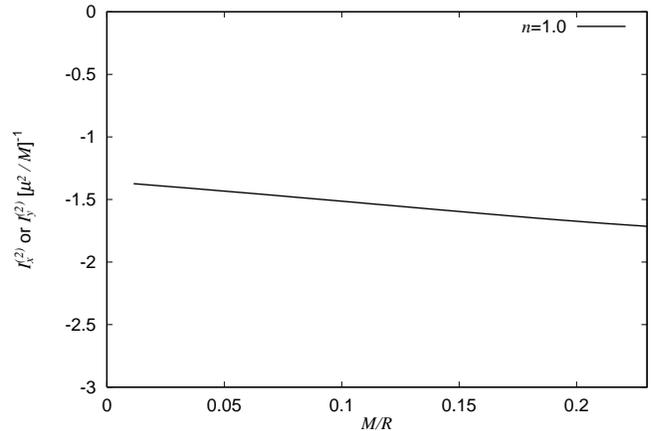}
  \vspace{0cm}
  \caption{The other components of the principal
     moments of inertia $I_{x}^{(2)}$ and $I_{y}^{(2)}$,
     which are derived for $n=1$.}
  \label{figix}
\end{figure}

However, there is no guarantee that all the results in 
Fig.~\ref{figiz} include monopole contributions
dominantly. If there is the case in which the monopole
part can be neglected, then we can estimate $I_{x}^{(2)}$
and $I_{y}^{(2)}$ in the case, by assuming that the relation
(\ref{q-rel}) holds approximately.
That case would also provide some extrapolation for 
general relativistic effects on $I_{x}^{(2)}$
and $I_{y}^{(2)}$ in the other cases.
In order to seek such a case, we now compare
$\Delta I_{1}$ derived by simply using the results
in Fig.~\ref{figiz} with $\Delta I_{2}$
calculated from the ellipticity, which was already
estimated for several stellar models
(Konno et al.~\cite{kok2}).
In the case that $\Delta I_{1}$ is almost consistent
with $\Delta I_{2}$ about the values and tendency,
it seems that the monopole part can be neglected,
and we may use the relation (\ref{q-rel}).
From the comparison, we find that $\Delta I_{1}$ is almost consistent
with $\Delta I_{2}$ in the case of $n=1$.
Figure \ref{figid} displays the comparison between 
$\Delta I_{1}$ and $\Delta I_{2}$ in this stellar model.
The two curves coincide within 10\%.
Hence, we can estimate $I_{x}^{(2)}$ and $I_{y}^{(2)}$ using 
Eq.~(\ref{q-rel}) in this stellar model.
Since the values of $I_{x}^{(2)}$ and $I_{y}^{(2)}$ are simply 
given by multiplying $I_{z}^{(2)}$ by a factor of $-1/2$,
the changes of the absolute values of $I_{x}^{(2)}$ and $I_{y}^{(2)}$ 
due to general relativistic effects
are specified by the same factor as in the case of $I_{z}^{(2)}$
(see Fig.~\ref{figix}).
Consequently, we derive very similar result to that of $I_{z}^{(2)}$.
However, $I_{x}^{(2)}$ and $I_{y}^{(2)}$ decrease with the
general relativistic factor $M/R$, while $I_{z}^{(2)}$ increases.
We expect that similar features exist also in the other
stellar models.

%

\section{Conclusions}
\label{con}

Recent observations indicate precession of some pulsars.
In order to understand the dynamics of these pulsars in detail,
we have to know the moments of inertia, which play a
crucial role in the Euler equation of motion.
Motivated by this consideration, we have discussed the 
general relativistic effect on the moments of inertia
of magnetically deformed stars. 
By developing the formulation by Konno et al.~(\cite{kok1}),
we have considered the slow rotation of magnetically 
deformed stars on the symmetric axis
in order to define the moment of inertia.
The general expression for the magnetic correction
$I_{z}^{(2)}$ to the moments of inertia of spherically
symmetric stars was obtained, and the numerical
estimates for various polytropic stellar models were
also obtained.
The other components of the moments of inertia
$I_{x}^{(2)}$ and $I_{y}^{(2)}$ were discussed
from the extension of relations in Newtonian gravity.
From those results, we found that  
each principal moment of inertia is modified by a factor
of 2 at most due to the general relativistic effect.
Nevertheless, it seems that the general relativistic effect
does not affect precessing motion noticeably in the case of
free precession of pulsars.
However, it is not clear whether or not the electromagnetic 
radiation reaction torque acting on the pulsars change
due to the general relativistic effect.
If this torque is modified by the general relativistic 
effect, then the period of the wobbles would also be
modified.
Therefore, this point should be made clear in future
investigation.

\begin{acknowledgements}
  I would like to thank Y.~Kojima for careful reading of the
  manuscript and for fruitful discussions and suggestions. 
  I would also like to thank M.~Kasai, H.~Asada and 
  K.~Ioka for valuable and useful comments, 
  and M.~Hosonuma for useful discussions.
  This work was supported in part by a Grant-in-Aid for 
  Scientific Research Fellowship of the Ministry of 
  Education, Science, Sports and Culture of Japan 
  (No.12001146).
\end{acknowledgements}

\appendix

\section{Summary of the perturbative approach}
\label{app}
 
we now review the perturbative approach 
developed by Konno et al.~(\cite{kok1}) and
Konno \& Kojima (\cite{kk})

\subsection{Equations of order $\varepsilon_{B}$}

First, we consider the magnetic field described by $a_{\phi}$, 
which has the lowest order. 
This function obeys the following equation
derived from the Maxwell equation,
\begin{equation}
\label{a-phi}
 e^{- \lambda} \frac{d^2 a_{\phi}}{dr^2}
 + \frac{1}{2} \left( \nu' - \lambda' \right)
  e^{- \lambda} \frac{d a_{\phi}}{dr}
 - \frac{2}{r^2} a_{\phi} = - 4\pi j_{\phi} .
\end{equation}
The electric current $j_{\phi}$ must satisfy the integrability 
condition (Ferraro \cite{ferraro}; Bonazzola et al.~\cite{bgsm})
\begin{equation}
 j_{\phi} = c_{0} r^2 \left( \rho_{0} + p_{0} \right) ,
\end{equation}
where $c_{0}$ is a constant of order $\varepsilon_{B}$.
This corresponds to the condition in which the star is in
equilibrium.
Equation (\ref{a-phi}) should be solved by imposing the condition
in which $a_{\phi}$ vanishes at the stellar center and infinity
and is connected smoothly at the stellar surface.
The exterior solution can be obtained analytically
(Ginzburg \& Ozerno{\u\i} \cite{go}; Petterson \cite{petter}),
while the interior solution was obtained numerically
(Konno et al.~\cite{kok1}).

\subsection{Equations of order $\varepsilon_{B}^2$}

Next, we consider the effect of magnetic stress, which
is of second order in $\varepsilon_{B}$. 
This effect arises from the electromagnetic part of the
stress-energy tensor (\ref{s-e-t}).
The second-order metric functions $h_{0}$, $h_{2}$, $m_{0}$,
$m_{2}$ and $k_{2}$ can be obtained by solving the following
two sets of differential equations and one algebraic equation
derived from the Einstein equation 
$G^{\mu}_{\ \nu} = 8\pi T^{\mu}_{\ \nu}$
and the equation of motion $T^{\mu}_{\ \nu ; \mu} = 0$,
\begin{eqnarray}
\label{de-m0}
 \frac{dm_{0}}{dr} & = & - 4\pi r^2 \frac{\rho'_{0}}{p'_{0}}
   \left( \rho_{0} + p_{0} \right) \left( h_{0} - c_{1} \right) 
   \nonumber \\
 & & + \frac{1}{3} e^{-\lambda} \left( a'_{\phi} \right)^2
  + \frac{2}{3r^2} a_{\phi}^2
  + \frac{8\pi}{3} \frac{\rho'_{0}}{p'_{0}} a_{\phi} j_{\phi} , \\
\label{de-h0}
 \frac{dh_{0}}{dr} & = & \left( \frac{1}{r^2} + \frac{\nu'}{r} \right)
  e^{\lambda} m_{0} - 4\pi  r e^{\lambda} 
  \left( \rho_{0} + p_{0} \right) \left( h_{0} - c_{1} \right)
  \nonumber \\
 & & + \frac{1}{3r} \left( a'_{\phi} \right)^2
  - \frac{2}{3r^3} e^{\lambda} a_{\phi}^2
  + \frac{8\pi}{3r} e^{\lambda} a_{\phi} j_{\phi} , 
\end{eqnarray}
\begin{eqnarray}
\label{de-v2}
 \frac{dv_{2}}{dr} & = & - \nu' h_{2} 
  + \frac{2}{3} e^{-\lambda} \left( \frac{1}{r}
    + \frac{\nu'}{2}\right)\left( a'_{\phi} \right)^2
  + \frac{4}{3r^2} a_{\phi} a'_{\phi} , \\
\label{de-h2}
 \frac{dh_{2}}{dr} & = & - \frac{4e^{\lambda}}{r^2 \nu'} v_{2}
  + \left[ 8\pi \frac{e^{\lambda}}{\nu'}
  \left( \rho_{0} + p_{0} \right) + 
  \frac{2\left( 1 - e^{\lambda} \right)}{r^2 \nu'}
  - \nu' \right] h_{2} 
  \nonumber \\ 
 & & + \frac{8}{3r^4 \nu'} e^{\lambda} a_{\phi}^2
    + \frac{8}{3r^3 \nu'} \left( 1+ \frac{r\nu'}{2} \right)
      a_{\phi} a'_{\phi} \nonumber \\
 & & + \left( \frac{1}{3} \nu' e^{-\lambda} 
  + \frac{2}{3r^2\nu'} \right) \left( a'_{\phi} \right)^2
  + \frac{16\pi}{3r^2\nu'} e^{\lambda} a_{\phi} j_{\phi} ,
\end{eqnarray}
\begin{equation}
 m_{2} = r e^{- \lambda} \left[ - h_{2} 
  + \frac{2}{3} e^{- \lambda} \left( a'_{\phi} \right)^2 \right] ,
\end{equation}
where $c_{1}$ is a constant fixed by the junction condition at the
stellar surface, and $v_{2}$ is defined as $v_{2} = h_{2} + k_{2}$.
The differential equations (\ref{de-m0})--(\ref{de-h2})
should be solved by imposing the boundary conditions
\begin{equation}
  m_{0}, h_{2}, k_{2} \rightarrow 0 , \quad 
   h_{0} \rightarrow \mbox{const} \quad 
   \mbox{as} \quad r \rightarrow 0 ,
\end{equation}
\begin{eqnarray}
\label{bc.metric.infty}
 \lefteqn{h_{0}, h_{2} \rightarrow 0 , \quad 
  m_{0} \rightarrow \mbox{const} , \quad 
  k_{2} \rightarrow \frac{1}{r^{\alpha}} \left( \alpha \ge 3 \right)} 
  \nonumber \\
 & & \mbox{as} \quad r \rightarrow \infty .
\end{eqnarray}
Here, note that the value of $m_{0}$ at infinity
corresponds to mass shift associated with the perturbation.
Furthermore, the junction condition at the stellar surface
is written as (O'Brien \& Synge \cite{os})
\begin{eqnarray}
\label{junc-1}
 \left. g_{\mu \nu} \right|_{+ \xi}
  & = & \left. g_{\mu \nu} \right|_{- \xi} \quad 
    \left( \mu , \nu = t, r, \theta, \phi  \right) , \\
\label{junc-2}
 \left. g_{\lambda \sigma , r} \right|_{+ \xi}
  & = & \left. g_{\lambda \sigma , r} \right|_{- \xi} \quad 
    \left( \lambda , \sigma = t, \theta, \phi  \right) ,
\end{eqnarray}
where $\xi$ denotes the boundary surface of the star.

Once we have the above metric functions, we can derive the 
second-order corrections of pressure and density from the relations
\begin{eqnarray}
 p_{20} & = & - \left( \rho_{0} + p_{0} \right) h_{0}
  + \frac{2}{3r^2} a_{\phi} j_{\phi}
  + c_{1} \left( \rho_{0} + p_{0} \right), \\
 p_{22} & = & - \left( \rho_{0} + p_{0} \right) h_{2}
  - \frac{2}{3r^2} a_{\phi} j_{\phi} , \\
 \rho_{20} & = & \frac{\rho'_{0}}{p'_{0}} p_{20} , \\
 \rho_{22} & = & \frac{\rho'_{0}}{p'_{0}} p_{22} .
\end{eqnarray}
Therefore, the stellar structure which is subject to 
magnetic deformation can be determined by solving 
Eqs.~(\ref{de-m0})--(\ref{de-h2}).
The exterior solution for the metric functions
was obtained analytically, while the interior solution
was calculated numerically by Konno et al.~(\cite{kok1}).
The fact that the evaluation of ellipticity using 
the results of the metric functions reproduced the Newtonian result
derived by Ferraro (\cite{ferraro}) in the Newtonian limit
(see Konno et al.~\cite{kok2}) justifies this formalism.

\subsection{Equations of order $\varepsilon_{\Omega} \varepsilon_{B}$}

Finally, We consider the quantities $a_{t0}$ and $a_{t2}$ of order 
$\varepsilon_{B} \varepsilon_{\Omega}$, which are related with
the electric fields induced by stellar rotation.
These functions obey the Maxwell equation. 
Outside the star, we have
\begin{eqnarray}
\label{eq.at0}
 \lefteqn{e^{-\lambda} \frac{d^2 a_{t0}}{dr^2}
   + \frac{2 e^{-\lambda}}{r} \frac{d a_{t0}}{d r}} \nonumber \\
 & & = \frac{2}{3} \left[ \left( \lambda' + \frac{2}{r} \right) \omega 
  + \omega' \right] e^{-\lambda} a'_{\phi}
  + \frac{4}{3r^2} \omega a_{\phi} , \\
\label{eq.at2}
 \lefteqn{e^{-\lambda} \frac{d^2 a_{t2}}{dr^2}
   + \frac{2 e^{-\lambda}}{r} \frac{d a_{t2}}{d r} 
   - \frac{6}{r^2} a_{t2}} \nonumber \\
 & & = -\frac{2}{3}  \left[ \left( \lambda' + \frac{2}{r} \right) \omega 
  + \omega' \right] e^{-\lambda} a'_{\phi}
  + \frac{8}{3r^2} \omega a_{\phi} , 
\end{eqnarray}
where the function $\omega$ of first order in $\varepsilon_{\Omega}$
was discussed in detail by Hartle (\cite{hartle}).
These differential equations can also be solved analytically 
(see Muslimov \& Tsygan \cite{mt}; Konno \& Kojima \cite{kk}
for the detailed calculations).
Inside the star, the assumption of a perfectly conducting interior
yields the expressions (Bonazzola et al.~\cite{bgsm}; 
Bocquet et al.~\cite{bbgn}),
\begin{eqnarray}
 a_{t0} & = & \frac{2}{3} \Omega a_{\phi} + c_{2} , \\
 a_{t2} & = & - \frac{2}{3} \Omega a_{\phi} ,
\end{eqnarray}
where $c_{2}$ is a constant fixed by the junction condition
at the stellar surface.
In this case, $a_{t0}$ and $a_{t2}$ are not connected
smoothly at the stellar surface, owing to the induced
surface charge.


\begin{thebibliography}{}

   \bibitem[1996]{bs}
      Baykal A., Swank J., 1996,
      ApJ 460, 470

   \bibitem[1998]{bsss}
      Baykal A., Swank J.H., Strohmayer T., Stark M.J., 1998,
      A\&A 336, 173

   \bibitem[1973]{ba}
      Bertotti B., Anile A.M., 1973,
      A\&A 28, 429.

   \bibitem[1995]{bbgn}
      Bocquet M., Bonazzola S., Gourgoulhon E., Novak J., 1995,
      A\&A 301, 757

   \bibitem[1993]{bgsm}
      Bonazzola S., Gourgoulhon E., Salgado M., Marck J.A., 1993,
      A\&A 278, 421

   \bibitem[1974]{cm}
      Chandrasekhar S., Miller J.C., 1974,
      MNRAS 167, 63

   \bibitem[2000]{cj}
      Cutler C., Jones D.I., 2000,
      preprint: gr-qc/0008021

   \bibitem[1995]{dmhd}
      D'Alessandro F., McCulloch P.M., Hamilton P.A., Deshpande A.A.,
      1995, MNRAS 277, 1033

   \bibitem[1970]{dg}
      Davis L., Goldstein M., 1970,
      ApJ 159, L81.

   \bibitem[1992]{dt}
      Duncan R.C., Thompson C., 1992,
      ApJ 392, L9

   \bibitem[1954]{ferraro}
      Ferraro V.C.A., 1954,
      ApJ 119, 407

   \bibitem[1970]{geroch}
      Geroch R., 1970
      J.~Math.~Phys.~11, 2580

   \bibitem[1965]{go}
      Ginzburg V.L., Ozerno{\u\i} L.M., 1965,
      Sov.~Phys.~JETP 20, 689

   \bibitem[1970]{goldreich}
      Goldreich P., 1970,
      ApJ 160, L11

   \bibitem[1974]{hansen}
      Hansen R.O., 1974,
      J.~Math.~Phys.~15, 46

   \bibitem[1967]{hartle}
      Hartle J.B., 1967,
      ApJ 150, 1005

   \bibitem[1999]{kh}
      Kojima Y., Hosonuma M., 1999,
      ApJ 520, 788

   \bibitem[2000]{kk}
      Konno K., Kojima Y., 2000,
      Prog.~Theor.~Phys., (in press)

   \bibitem[1999]{kok1}
      Konno K., Obata T., Kojima Y., 1999,
      A\&A 352, 211

   \bibitem[2000]{kok2}
      Konno K., Obata T., Kojima Y., 2000,
      A\&A 356, 234

   \bibitem[1998]{sgr1}
      Kouveliotou C., Dieters S., Strohmayer T., et al., 1998,
      Nat 393, 235

   \bibitem[1999]{sgr2}
      Kouveliotou C., Strohmayer T., Hurley K., et al., 1999,
      ApJ 510, L115

   \bibitem[1999]{melatos1}
      Melatos A., 1999,
      ApJ 519, L77

   \bibitem[2000]{melatos2}
      Melatos A., 2000,
      MNRAS 313, 217

   \bibitem[1995]{mereghetti}
      Mereghetti S., 1995,
      ApJ 455, 598

   \bibitem[1995]{ms}
      Mereghetti S., Stela L., 1995,
      ApJ 442, L17

   \bibitem[1992]{meszaros}
      M\'esz\'aros P., 1992,
      High-Energy Radiation from Magnetized Neutron Stars,
      The University of Chicago Press, USA

   \bibitem[1986]{mt}
      Muslimov A.G., Tsygan A.I., 1986,
      SvA 30, 567.

   \bibitem[1952]{os}
      O'Brien S., Synge J.L., 1952,
      Comm.~Dublin Inst.~Advanced Studies, A no.~9

    \bibitem[1998]{opmi}
      Oosterbroek T., Parmar A.N., Mereghetti S., Israel G.L., 1998,
      A\&A 334, 925

    \bibitem[1974]{petter}
      Petterson J.A., 1974
      Phys.~Rev.~D10, 3166

    \bibitem[1976]{quin}
      Quintana H., 1976,
      ApJ 207, 279  
 
    \bibitem[1994]{rp}
      Ravenhall D.G., Pethick C.J., 1994,
      ApJ 424, 846

    \bibitem[1996]{glitch1}
      Shemar S.L., Lyne A.G., 1996,
      MNRAS 282, 677

    \bibitem[2000]{sls}
      Stairs I.H., Lyne A.G., Shemar S.L., 2000,
      Nat 406, 484

    \bibitem[1993]{td1}
      Thompson C., Duncan R.C., 1993,
      ApJ 408, 194

    \bibitem[1995]{td2}
      Thompson C., Duncan R.C., 1995,
      MNRAS 275, 255

    \bibitem[1996]{td3}
      Thompson C., Duncan R.C., 1996,
      ApJ 473, 322

    \bibitem[2000]{glitch2}
      Wang N., Manchester R.N., Pace R.T., et al., 2000,
      MNRAS 317, 843
\end{thebibliography}
\end{document}